\documentclass{vgtc}                          % final (conference style)
\ifpdf%                                % if we use pdflatex
  \pdfoutput=1\relax                   % create PDFs from pdfLaTeX
  \usepackage{graphicx}                % allow us to embed graphics files
  \DeclareGraphicsExtensions{.pdf,.png,.jpg,.jpeg} % for pdflatex we expect .pdf, .png, or .jpg files
\else%                                 % else we use pure latex
  \usepackage{graphicx}                % allow us to embed graphics files
  \DeclareGraphicsExtensions{.eps}     % for pure latex we expect eps files
\fi%

\graphicspath{{figures/}{pictures/}{images/}{./}} % where to search for the images

\usepackage{microtype}                 % use micro-typography (slightly more compact, better to read)
\PassOptionsToPackage{warn}{textcomp}  % to address font issues with \textrightarrow
\usepackage{textcomp}                  % use better special symbols
\usepackage{mathptmx}                  % use matching math font
\usepackage{times}                     % we use Times as the main font
\usepackage{cite}                      % needed to automatically sort the references
\usepackage{tabu}                      % only used for the table example
\usepackage{booktabs}                  % only used for the table example
\usepackage{flushend}

\usepackage{color}
\usepackage{xcolor}

\onlineid{0}

\vgtccategory{Research}

\vgtcinsertpkg

\title{The Huge Variable Space in Empirical Studies for Visualization ---\\A Challenge as well as an opportunity for Visualization Psychology}

\author{Min Chen\thanks{e-mail: min.chen@oerc.ox.ac.uk}\\ %
        \scriptsize University of Oxford, UK %
\and Alfie Abdul-Rahman\thanks{e-mail: alfie.abdulrahman@kcl.ac.uk}\\ %
     \scriptsize King's College London, UK %
\and David H. Laidlaw\thanks{e-mail: dhl@cs.brown.edu}\\ %
     \scriptsize Brown University, USA}

\abstract{In each of the last five years, a few dozen empirical studies appeared in visualization journals and conferences.
The existing empirical studies have already featured a large number of variables.
There are many more variables yet to be studied. 
While empirical studies enable us to obtain knowledge and insight about visualization processes through observation and analysis of user experience, it seems to be a stupendous challenge for exploring such a huge variable space at the current pace.
In this position paper, we discuss the implication of not being able to explore this space effectively and efficiently, and propose means for addressing this challenge.

} % end of abstract

\begin{document}


\firstsection{Introduction}

\maketitle

%% \section{Introduction} %for journal use above \firstsection{..} instead

Many controlled and semi-controlled empirical studies have provided empirical evidence to compare and measure effectiveness and efficiency of different visualization  techniques (or approaches, algorithms, systems, workflows, and so on).
Some have provided support to existing theories or models for visualization and visual analytics, while several have challenged some commonly-known assumptions, wisdoms, and guidelines.
Most of such studies consist of one or a few experiments, each features a few independent and dependent variables.
One might wish for empirical studies to capture all possible independent variables that may be featured in commonly-used visual representations and all dependent variables that could be used to measure the performance of typical visualization tasks.
However, the sheer number of these variables present a hindrance to any controlled or semi-controlled studies.
On the other hand, distributing these variables to many studies, each focusing on a few variables, demands a large research community and a lot of resources.

Recently Abdul-Rahman et al. conducted a survey of 32 empirical study papers \cite{Abdul-Rahman:2019:survey}.
They identified some 64 types of independent variables, and categorized them into five classes.
The first four classes (56 types) all focused on visual signals, while the fifth class (8 types) focused on non-visual variables (e.g., task, teaching method, etc.).
%
% \begin{enumerate}
%    \vspace{-2mm}
%    \item Varying values in a single visual channel or varying types of visual channels;
%    \vspace{-2mm}
%    \item Varying visual objects featuring multiple visual channels or the characteristic attributes of the combined variations;
%    \vspace{-2mm}
%    \item Varying visual patterns made of multiple visual objects or the characteristic attributes of the visual patterns;
%    \vspace{-2mm}
%    \item Varying plot types or plot-level visual designs;
%    \vspace{-2mm}
%    \item Varying variables not in the depicted data.
% \end{enumerate}
%
They observed that ``\emph{there is no shortage of studies on independent variables in each category},'' but ``\emph{there are many more research questions yet to be asked or answered, and the scope of visualization-related empirical studies is huge}.''
They concluded:

\begin{itemize}
    \vspace{-2mm}
    \item[] ``\emph{It may thus be desirable for the visualization researchers who conduct empirical studies to be more coherently organized, instead of being distributed sparsely in InfoVis, SciVis, VAST, and other areas of visualization. This will allow these researchers to share their expertise (e.g., in the review processes) more easily and to formulate research agenda in a more ambitious and structured manner.}'' ``\emph{By providing some opportunities to bring all these researchers together, we may soon see the emergence of a new area of \textbf{visualization psychology}.}''
    \vspace{-1mm}
\end{itemize}

\vspace{-1mm}
\noindent This echoes an earlier observation in another survey \cite{Abdul-Rahman:2019:arXiv}: ``\emph{There  are  many  branches  of applied  psychology ... One has to ask that `is there a  room  for \textbf{visualization  psychology}?'} ''
In this position article, we provide further discourse on how to address the huge variable space in visualization psychology.

\section{Observations}
The main obstacles to the scalability of empirical studies in visualization include
(i) the relatively small number of visualization researchers who design and conduct empirical studies,
(ii) the complex variations in visualization in a combinatoric manner, and
(iii) the narrow hypothesis-based experimental design suitable for publication requirements. A new area of visualization psychology may adopt the following strategies to help overcome these obstacles.

\vspace{2mm}\noindent\textbf{More Experimental Scientists.}
Building on the references collected by Lam et al. \cite{Lam:2012:TVCG}, Kijmongkolchai et al. \cite{Kijmongkolchai:CGF:2017}, Fuchs et al. \cite{Fuchs:2016:TVCG}, and Roth et al.\cite{Roth:2017:IJC}, Abdul-Rahman et al. surveyed 129 papers on visualization-focused empirical studies \cite{Abdul-Rahman:2019:arXiv} until 2018. Their statistics show that on average the Journal of Psychological Review published about 38 papers per year between 1978 and 2018, while the average number of visualization-focused empirical studies is about 12 per year between 2010 and 2018.
Considering that a Wikipedia page lists 144 psychology journals, the empirical studies that focus on visualization and visual analytics are drops in the ocean.
The situation is unlikely to improve substantially within the field of visualization as the overall number of scientists, researchers, and practitioners is small, while a large portion of them are busy with other subareas, such as  applications, systems, algorithms, designs, theories, and so on. 
Having Visualization Psychology as an interdisciplinary field and a branch of applied psychology can potentially attract many researchers in psychology to design and conduct experiments focused on or closely related to visualization.

\vspace{2mm}\noindent\textbf{More Studies on the ``Mind''.}
Most visualization-focused empirical studies examine hypotheses about the \emph{artefacts} in visualization images.
For example, Laidlaw et al. compared four techniques for visualizing 2D vector fields \cite{Laidlaw:2001:Vis}, Chen et al. compared four visual representations for depicting motion signatures in videos \cite{Chen:2006:TVCG}, and Kanjanabose et al. compared data tables, scatter plots and parallel coordinates plots \cite{Kanjanabose:2015:CGF}.
Sometimes, such studies of artefacts (e.g., techniques, plots, visual representations, systems, etc.) have led to findings about the \emph{mind}.
In their artefact-based study, Chen et al. \cite{Chen:2006:TVCG} by chance discovered that participants unconsciously remembered the video visualization skills acquired in the first study and performed better three months later in the second study than those who did not take part in the first study. This is a finding about memory and learning -- aspects of cognition. 
Similarly, when studying data tables, scatter plots and parallel coordinates plots, Kanjanabose et al. \cite{Kanjanabose:2015:CGF} found that participants could retrieve data values more quickly and accurately with data tables than with scatter plots and parallel coordinate plots. Since visualization was commonly considered as a means for viewing data values, and many empirical studies compare artefacts with data retrieval tasks, this raises a question: what would have happened if data tables had been involved in the comparison, or more fundamentally, in what condition visualization is better for data retrieval tasks than data tables?

In recent years, more studies were designed explicitly to study the mind, and artefacts were moved to a secondary role as stimuli for observing the mind. There have been studies on memory \cite{Boyandin:2012:CGF}, attention \cite{Haroz:2012:TVCG}, visual grouping \cite{Gramazio:2014:TVCG}, knowledge \cite{Kijmongkolchai:CGF:2017}, and so on. Although artefacts were used as stimuli, the experimenters were aiming for discoveries about the mind, which can be applied to other artefacts that were not examined in the studies. For example, when Szafir found that the perception of colors was size-dependent \cite{Szafir:2018:TVCG}, this naturally led to many hypotheses that the perception of \textbf{A} might be \textbf{Y}-dependent. It could also lead to more fundamental hypotheses: must visual encoding be always isomorphic and can it be polymorphic \cite{Chen:2020:book} since human perception could not hold up the isomorphic requirement anyway \cite{Szafir:2018:TVCG}? If the latter is true, what cognitive factor may condition polymorphic perception?

Focusing on the mind potentially allows empirical research in visualization to make a big stride in making fundamental advances in the field of visualization. It is likely that studying the mind is hard than studying artefacts. However, any discovery about the mind can be translated to inferences about many artefacts. Of course, this is not to say that we should not study artefacts. Indeed, as mentioned earlier, findings about artefacts can lead to hypotheses and potentially major discoveries about the mind. Building on the past studies of artefacts, empirical researchers in visualization, hopefully, together with more and more colleagues in psychology, we will be able to conduct more studies on the mind.
 
\vspace{2mm}\noindent\textbf{More Progressive Approaches.}
Studying a hypothesis about the mind is entrenched in almost every empirical study in psychology. It is also a tradition in psychology that a hypothesis typically investigated in many empirical studies by several teams. It has been rare that a hypothesis is confirmed or disproved after the first empirical study on the hypothesis.
A switching of emphasis from artefacts to the mind may instigate more progressive approaches to studying a challenging hypothesis. 

Firstly, empirical researchers in visualization should embrace the tradition of psychology in scholarly contention and disagreement, and should welcome any serious challenge to an existing theory or finding as long as there is an adequate empirical evidence or analytical rationale suggesting that the existing theory or finding might not be 100\% correct as many thought. While it is not easy for reviewers to read papers that challenge their past theories or findings, reviewers in such situation should exercise a high level of integrity and professionalism, e.g., in making an objective assessment, declaring a conflict of interest if appropriate, and overcoming the prepossession for suppressing the debate through nitpicking.

Secondly, empirical researchers in visualization may explore other forms of empirical studies that do not involve controlled or semi-controlled experiments.
The BELIV Workshop (\url{https://beliv-workshop.github.io/}) is a biennial event. Since it was established in 2006, it has been encouraging empirical researchers to develop ``\emph{new and innovative evaluation methods for visualization tools and techniques.}''
While BELIV has a strong focus on artefacts, findings obtained from the evaluation of some visual representations, interaction techniques, and visualization tools can also inform the development of new hypotheses, conceptual models, and qualitative theories about the mind in the context of visualization.

Thirdly, visualization scientists are data scientists and are used to process a variety of data using data mining and data visualization.
Meanwhile, empirical studies, controlled as well as uncontrolled, collect data about various variables in visualization processes, including the variables about artefacts as well as those about the mind. 
Often such data may not be adequate for confirming a binary hypothesis in a statistically significant manner.
It may feature too many variables, or some variables may have too many values that cannot be clustered into a few groups. 
Nevertheless, if the collected data features some strong variations in the relation between the independent and dependent variables, we can discover such relations using visual analytics workflows where statistics, algorithms, visualization, and interaction are integrated.
We can also use the data to develop data-driven models and data-driven metrics.
Such a model or metric defines a complex causal relation in a probabilistic or functional manner, which is sometimes perceived to be less definite than a hypothesis confirmed by an empirical study. In fact, a data-driven model or a data-driven metric is just an intermediate step stone towards a grand theory.
Empirical researchers in visualization should welcome and embrace such intermediate steps, simply because studying a hypothesis about the mind is usually much more complex than studying artefacts.
Evaluating whether artefact \textbf{A} is better than artefact \textbf{B} may need one or a few empirical studies.
Determining whether a function of the mind, \textbf{\emph{X}()}, causes \textbf{A} to be better than artefact \textbf{B} will likely require many intermediate steps.

\section{Conclusions}
In data science, interactive visualization and visual analytics brings together machine-centric processes and human-centric processes.
It can provide psychologists with one of the best platforms for studying the human mind.
Therefore, creating a new interdisciplinary area of visualization psychology  will not only benefit the research and development in the field of visualization, but also benefit the scientific agenda in psychology.
In particular, the aforementioned fundamental questions in visualization are also fundamental questions in perception and cognition.
Many currently imperfect guidelines in visualization reflect some limited understanding in terms of perception and cognition.
Failures or shortcomings in the human mind often inspire some best research topics in psychology.
Similarly, failures or shortcomings of visualization guidelines could inspire some best research topics in visualization psychology.

%In comparison with the discipline of psychology, the variables that have been studied in so far visualization-related empirical studies are only drops in the ocean \cite{Abdul-Rahman:2019:arXiv}. There is a huge number of variables that may represent different visual channels, visual patterns, visualization plots, visualization techniques, and factors that may affect human perception and cognition during visualization. With the current number of empirical researchers in the field of visualization, exploring such a huge variable space would take decades and centuries. Hence developing \textbf{\emph{Visualization Psychology}} as a new interdisciplinary research area as well as a new applied branch of psychology may attract more colleagues in psychology to study the mind in visualization processes.

Meanwhile, many visualization scientists and researchers are highly skilled in data analysis and have access to many practical applications. Visualization psychology can benefit from such skills and applications in developing new research methodologies and delivering high impact applications. 

Having more studies on the mind and having more progressive approaches naturally lead to an update of the existing evaluation criteria for artefact-focused empirical study papers. An accepted empirical study paper in visualization psychology may feature one of the following qualities:

\begin{itemize}
    \vspace{-2mm}
    \item \textbf{Novelty.}
    An empirical study reports new discoveries and findings that have not been previously obtained. The study may examine a new phenomenon in visualization, or provide evidence to support or contradict an unsupported theoretical hypothesis or practical wisdom.
    \vspace{-2mm}
    \item \textbf{Innovation.}
    An empirical study features new study methodologies that are previously unknown to or uncommon in visualization research, and are technically sound and beneficial in the direct and indirect observation of user experience and the collection of empirical data. Such a methodology may become a new template for empirical studies in visualization.
    \vspace{-2mm}
    \item \textbf{Significance.}
    An empirical study presents an experiment that is substantially more comprehensive, or leads to more meaningful statistical inference, than previous studies on the topic.%
    \vspace{-2mm}
    \item \textbf{Impact.} An empirical study that may lead to a significant change of our fundamental understanding about visualization, or result in new guidelines and practices in visualization. Such impact may have been evidently confirmed, or an initial assessment may have convincingly suggested the potential.
    \vspace{-2mm}
    \item \textbf{Data, Evidence, Measurement, and Analysis.}
    An empirical study reports important data samples, evidence, measurement, and analysis that have not been previously obtained. The study may contribute towards the discoveries and findings of a major, fundamental, and complex hypothesis that is difficult to confirm or disapprove through one or a few empirical studies.
\end{itemize}


%\bibliographystyle{abbrv}
\bibliographystyle{abbrv-doi}

\bibliography{references}
\end{document}